# SOT-MRAM-Enabled Probabilistic Binary Neural Networks for Noise-Tolerant and Fast Training

Puyang Huang[1,#], Yu Gu[1,#], Chenyi Fu[2,#], Jiaqi Lu[2], Yiyao Zhu[1], Renhe Chen[1], Yongqi Hu[1], Yi Ding[1], Hongchao Zhang[2], Shiyang Lu[2], Shouzhong Peng[2,*], Weisheng Zhao[2], and Xufeng Kou[1,*]
[1]ShanghaiTech University, Shanghai, China, [2]Beihang University, Beijing, China
*Email: kouxf@shanghaitech.edu.cn; shouzhong.peng@buaa.edu.cn , #Authors contributed equally to this work.

*Abstract*—We report the use of spin-orbit torque (SOT) magnetoresistive random-access memory (MRAM) to implement a probabilistic binary neural network (PBNN) for resource-saving applications. The in-plane magnetized SOT (i-SOT) MRAM not only enables field-free magnetization switching with high endurance ($> 10^{11}$), but also hosts multiple stable probabilistic states with a low device-to-device variation ($< 6.35\%$). Accordingly, the proposed PBNN outperforms other neural networks by achieving an 18× increase in training speed, while maintaining an accuracy above 97% under the write and read noise perturbations. Furthermore, by applying the binarization process with an additional SOT-MRAM dummy module, we demonstrate an on-chip MNIST inference performance close to the ideal baseline using our SOT-PBNN hardware.

## I. INTRODUCTION

With the advent of artificial intelligence, a seismic shift is observed in computing paradigms. As we move towards handling larger volumes of data and higher task complexities, new architectures for artificial neural networks (ANNs) have sprung up in numerous applications [1]. Conventionally, ANNs have relied on high-precision floating-point arithmetic to obtain optimal computational results. However, these approaches always require substantial computing and memory resources. Therefore, as the number of operations and data volume increase, the training process inevitably slows. In addition, the performance of these deterministic networks is heavily affected by discrepancies between standard training datasets and actual input data, which invariably result in reduced accuracy [2]. Alternatively, probabilistic-featured PBNNs, which introduce stochasticity within the network to enhance robustness, have been proposed to facilitate the convergence to global optima [3]. Consequently, the noise-tolerant PBNNs could offer accelerated training, robustness, and resource-saving capabilities for image/video classification and natural language processing.

In principle, the key of PBNNs relies on the conversion of probabilities into deterministic outcomes via data sampling [4]. This necessitates that the PBNN hardware withstands a large number of sampling operations, while the power consumption for each operation needs to be as low as possible. In this context, the non-destructive electrical manipulation of magnetic moments inherently allows MRAM to possess high endurance and energy-efficient write/read characteristics. More importantly, the magnetization switching probability can be well-controlled by the injection current level via spin-orbit torque, therefore making MRAM a suitable building block to construct PBNNs.

Inspired by the above scenario (Fig. 1), we utilize the SOT-MRAM platform to harness the PBNN advantages. The device across the 8-inch wafer exhibits highly consistent performance in terms of low resistance variation and identical probabilistic switching curves with repeatable state variables. By implementing both the vector-matrix multiplication (VMM) and the binarization operations with SOT-MRAM, we demonstrate a noise-tolerant and fast-training PBNN with an on-chip MNIST digit recognition accuracy of 90%.

## II. FIELD-FREE PROBABILISTIC SWITCHING OF IN-PLANE MAGNETIZED SOT-MRAM

### A. Device characterizations and SOT-driven probabilistic magnetization switching

High-quality magnetic tunnel junction (MTJ)/heavy metal (HM) thin films were prepared on 8-inch Si/SiO$_2$ wafers by magnetron sputtering. To enable field-free operation, we adopted the i-SOT MRAM configuration, in which an elliptical-shaped MTJ design warrants in-plane magnetic anisotropy. Accordingly, the spin current generated in the HM layer is parallel to the magnetization direction ($M$) of the free layer; therefore, the SOT is inserted to directly switch $M$ without the presence of an assisted magnetic field (Fig. 2). After film growth, large-scale SOT-MRAM array was fabricated, with typical device size of 0.7 μm × 2 μm. High-resolution transmission electron microscope (HR-TEM) images in Fig. 3 visualize the sharp hetero-interfaces. Subsequently, reliable field-free SOT-driven magnetization switching was demonstrated (Fig. 4), where the response time of the SOT-MRAM is below 400 ps, and the hysteresis window (*i.e.*, switching voltage of 400 ps pulse is $V_C$) of the $R$-$V$ curve is modulated by the pulse width. Moreover, because of the non-destructive SOT switching mechanism, the recorded parallel resistance ($R_P$) and antiparallel resistance ($R_{AP}$) in Fig. 5 did not experience any distortion after $10^{11}$ write/read cycles (*i.e.*, the tunneling magnetoresistance ratio of ~70% is sufficient for the VMM operation in PBNNs). Besides, by changing the input voltage around $V_0$ (voltage of 50% switching probability), multiple switching probability states (*i.e.*, network weights) are obtained (Fig. 6), and their corresponding probabilities are repeatable (*i.e.*, whose values were deduced by counting the number of $R_P$-to-$R_{AP}$ switching during 500 samplings per voltage).

Apart from single SOT device characterizations, device-to-device variations also play an important role in determining the overall network functionality. In this regard, Fig. 7 confirms that 8 randomly selected SOT-MRAM devices from the array all yielded 11 well-defined intermediate probabilistic states, with an average variation of 6.35% in the examined operating range. In the meantime, the standard deviation of the $R_P$ ($R_{AP}$) normal distribution curve, which was collected from 100+ devices across the 8-inch wafer, is found to be 4.72% (4.79%), as shown in Fig. 8. Besides, we need to point out that such a small resistance variation has a negligible impact on the MNIST classification accuracy of the subsequent simulations and PBNN on-chip validation. Consequently, the proposed SOT-MRAM provides stable and uniform probabilistic switching states, thereby laying a solid foundation for the design and implementation of PBNN.

### III. SOT-MRAM-ENABLED PBNN IMPLEMENTATION

*B. PBNN network structure and process flow for MNIST test*

To demonstrate the SOT-MRAM-enabled PBNN, we developed a network that consists of two convolutional layers, two max-pooling layers, and three fully connected layers (*i.e.*, all layers are constructed by SOT-MRAM) for standard MNIST handwritten digit recognition test (Fig. 9). Utilizing the faster training speed of PBNN, our PyTorch simulation results in Fig. 10 show that the ideal classification accuracy of PBNN quickly exceeds 98% after only 4 epochs, whereas other networks require at least 20 epochs under full-precision conditions [5-7]. Equivalently, the PBNN system can realize a significant training time reduction of 6× to 18× (Fig. 11). Another advantage of PBNN is its resource-saving feature. For instance, the entire network needs only eight quantized states for both weights and activations to achieve an accuracy above 98% at 30 epochs (Fig. 12). Furthermore, the PBNN displays a salient noise-tolerant property against the write and read errors. According to the error-awareness simulation data in Fig. 13, it is seen that the training result remains almost constant with respect to the write error, which may benefit from the natural stochasticity of probabilistic switching. On the other hand, even though the increase of the read error lowers the classification accuracy of PBNN to 90%, its performance is still better than that of other neural network counterparts [8]. Considering that the measured write and read errors of the SOT-MRAM devices are less than 6.35% (Fig. 7) and 4.8% (Fig. 8) respectively, the overall accuracy of our SOT-PBNN can reach the 97.33% benchmark in the ideal scenario.

*C. Hardware implementation of SOT-MRAM PBNN*

Guided by the PyTorch simulation, we further designed an on-chip PBNN system based on the in-plane magnetized SOT-MRAM array. As illustrated in Fig. 14, the row devices are selected through an SWL decoder, while the column devices are selected by write and read voltages from a digital-to-analog converter (DAC). Afterwards, a transimpedance amplifier (TIA) converts the current accumulated during the VMM operation into a voltage signal, which is binarized before passing to the next network layer. Given that the SOT-MRAM resistance changes with the read voltage, the binarization process in our SOT-PBNN system cannot be performed using a fixed-value resistor. Instead, to enable on-chip current comparison, we allocated an additional [2 × n] dummy cell corresponding to the [m × n] MRAM array cell. As a result, the binarization is achieved by writing $R_{AP}$ and $R_P$ into two MRAM devices along the same column of the dummy cell, and then determining the resistance state of a single MRAM using half of the summed currents under the same read voltage. The overview of the integrated SOT-MRAM PBNN chip is shown in Fig. 15.

As a proof-of-concept, the VMM operation was validated experimentally in a [16 × 1] array. Specifically, after all 16 serial-connected SOT-MARM devices were initialized in the anti-parallel state (*i.e.*, weight assignment), the accumulated current ($I_{out}$) measured at the output port is highly consistent with the ideal value ($I_{sum} = \sum_{i=1:16} I_{M0,i}$) under different read voltages (Fig. 16). It is also noted that with a maximum read voltage of 0.54 V, the average output current variation is only 4.31%, again demonstrating the low read error of our SOT-MRAM devices. Concurrently, the on-chip comparator function was evaluated from 9 randomly selected devices. Although the difference between $R_P$ and $R_{AP}$ becomes narrower as the read voltage increases, the reference resistance $R_{ref}$ of the dummy cell is continuously kept within the resistance gap, hence verifying a wide operating range of the SOT-MRAM dummy cell (Fig. 17). Finally, we selected an 8-level activation quantization and 3-level weight quantization for handwritten digit recognition. After transferring the weight values to MRAM, we conducted inference using the MNIST dataset. From the measured and simulated results in Fig. 18, it is seen that the inference currents distributions share the same feature versus the assigned weight information, therefore yielding the correct digit judgement. Based on the inference results after 100 training sessions, we have obtained an on-chip classification accuracy over 90% in our integrated SOT-PBNN chip (Fig. 19).

### IV. CONCLUSION

Compared with other memristive devices and neural networks, the SOT-MRAM-enabled PBNN elaborated in this work shows advantages including long endurance, stable states, fast training, and robustness against input variations (Table 1). Our work provides a compelling framework for the design of reliable neural networks for low-power applications with limited computational resources.

ACKNOWLEDGMENT

This work is supported by the National Key R&D Program of China (2021YFA0715503), the NSFC Programs (11904230, 62004013), the Shanghai Rising-Star Program (21QA1406000), and the Young Elite Scientists Sponsorship Program by CAST (2021QNRC001).

## Device structure characteristics and SOT switching performance

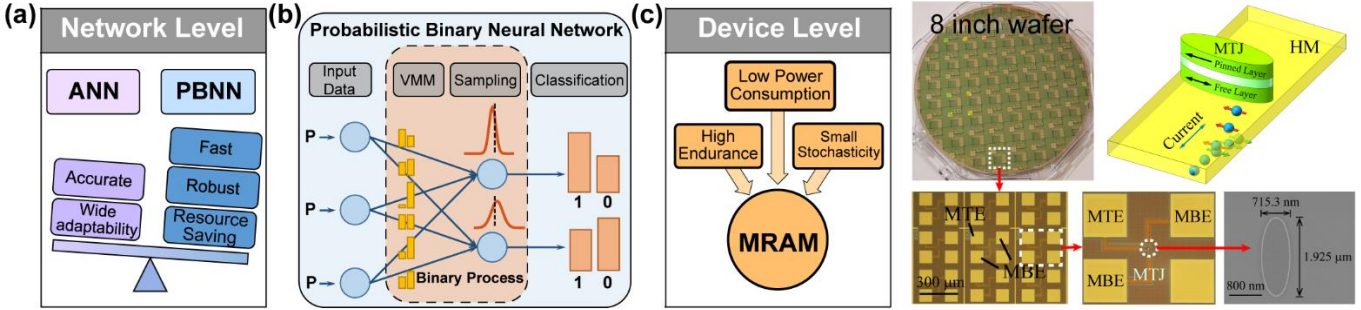

Fig. 1. (a) Illustration of the advantages of PBNN over conventional ANNs. (b) Operational diagram of PBNN training scheme with probabilistic and binary weights. (c) MRAM as an appealing candidate for PBNN implementation.

Fig. 2. Optical and SEM images of the 8-inch SOT-MRAM array with elliptical-shaped MTJ for field-free i-SOT switching.

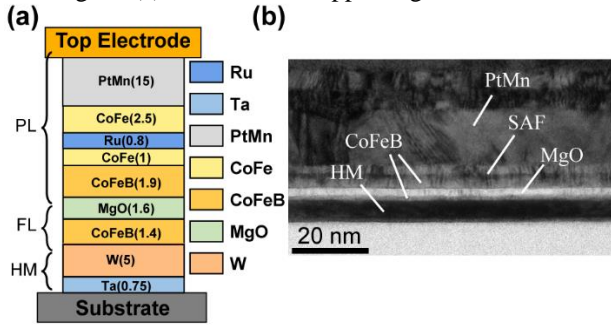
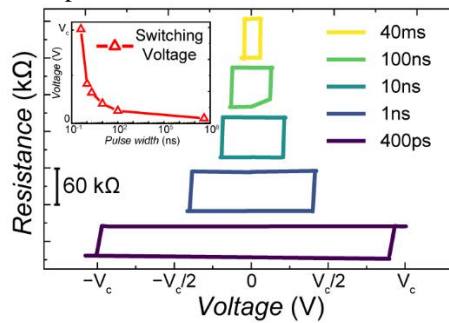
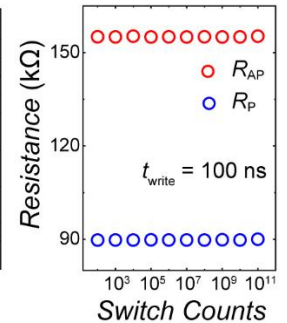

Fig. 3. (a) Film stack and (b) cross-sectional HR-TEM images of the fabricated i-SOT MRAM device with sharp film interfaces.

Fig. 4. SOT-driven switching curves with varied pulse width. Inset: Relation of the pulse width and switching voltage.

Fig. 5. MTJ $R_P$ and $R_{AP}$ experience no degradation after $10^{11}$ switching cycles.

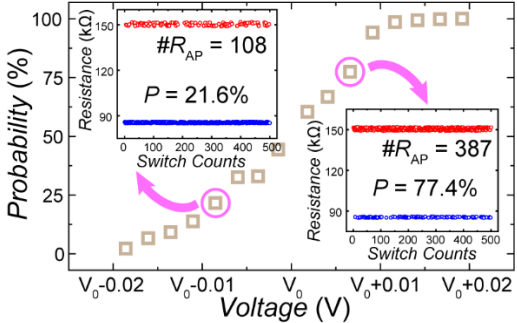
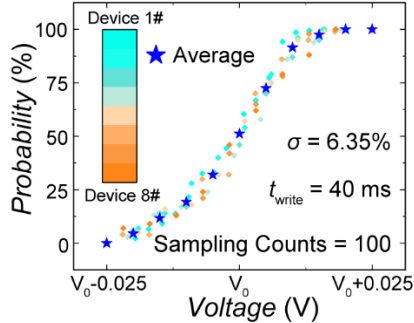
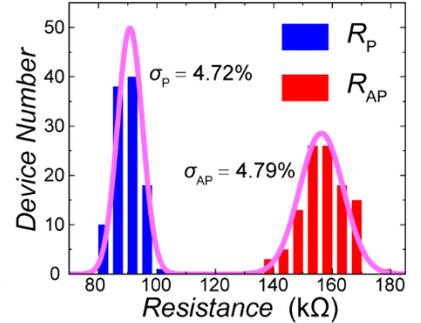

Fig. 6. Switching probability under various write voltages. Inset: P-to-AP switching distribution of each state, extracted from 500 sampling counts.

Fig. 7. Uniform probabilistic switching curves with 11 states taken from eight devices after 100 sampling counts.

Fig. 8. Normal distributions of the parallel ($R_P$) and antiparallel ($R_{AP}$) states of the SOT-MRAM array.

## PBNN Processing Flow and Conceptional Validation

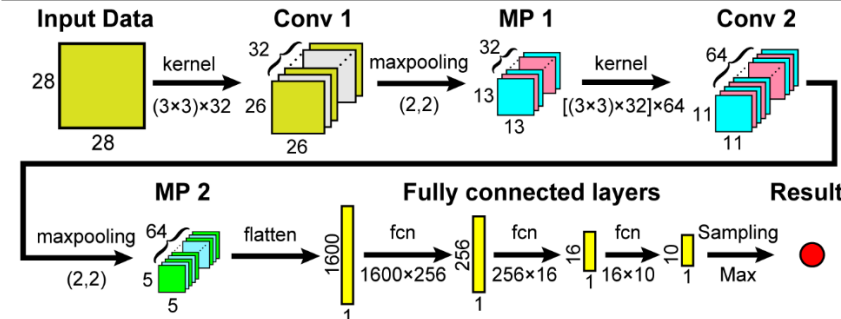
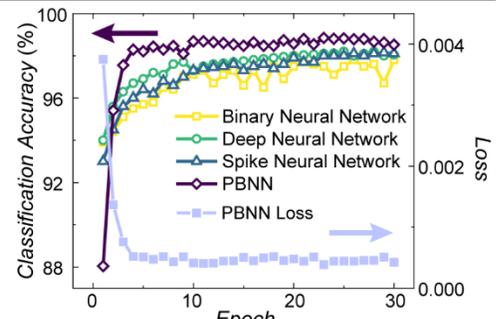

Fig. 9. The network structure for PBNN with 2 convolutional, 2 max-pooling and 3 fully-connected layers.

Fig. 10. Comparison of the classification accuracy among different neural networks.

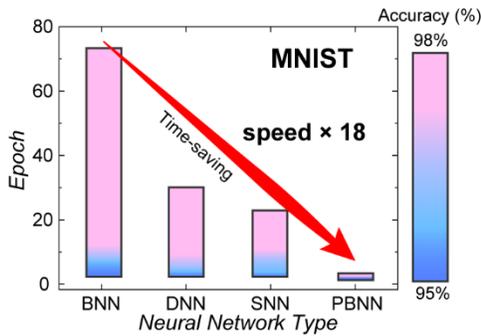
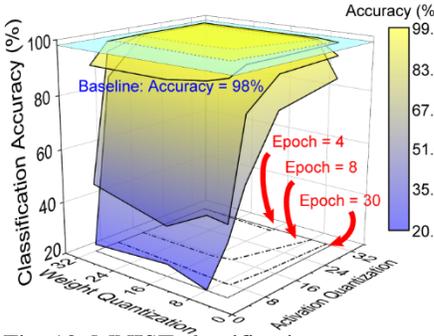
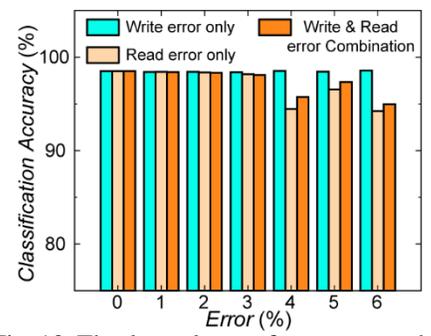

Fig. 11. Required epoch numbers for different neural networks to achieve reliable MNIST classification results.

Fig. 12. MNIST classification accuracy mapping versus weight quantization and activation quantization.

Fig. 13. The dependence of accuracy on the device switching probability variation during the write and read process.

## Hardware Orientated PBNN Design Scheme and Array Implementation

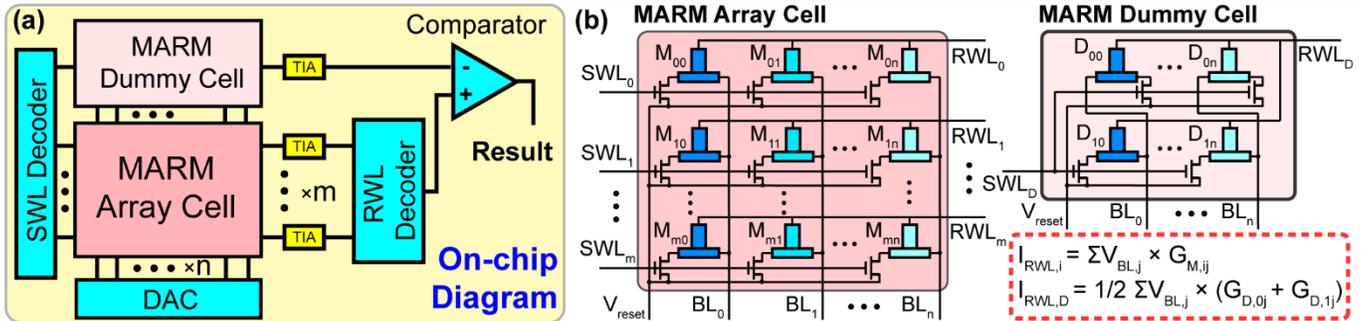

Fig. 14. (a) Schematic of the SOT-MRAM-enabled PBNN system, and (b) the circuit architectures of [m ×n] MRAM array cell (left panel) for vector-matrix multiplication process and [2 ×n] MRAM dummy cell (right panel) for binarization process.

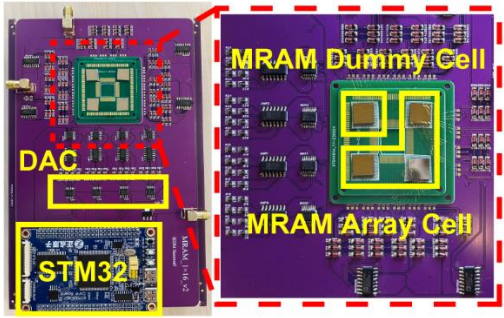
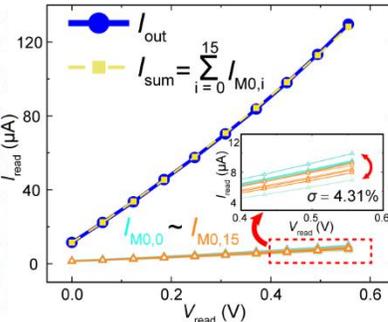
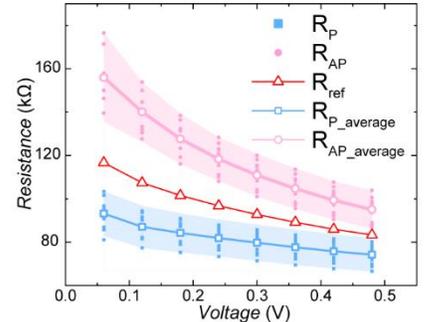

Fig. 15. The integrated SOT-MRAM PBNN chip and MNIST testing board setup.

Fig. 16. Measured output current $I_{out}$ after the VMM operation with 16 serial-connected devices. $I_{sum}$ is the ideal value.

Fig. 17. Voltage-dependent $R_P$ and $R_{AP}$ distributions of 9 devices. $R_{ref}$ is the reference resistance of the dummy cell.

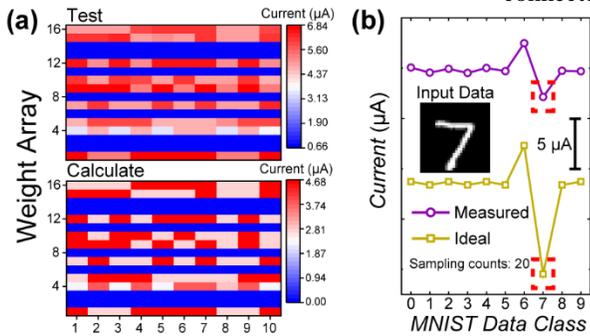
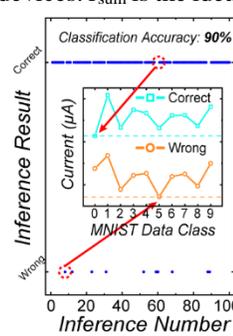

| | Device Level | | | | |
|---|---|---|---|---|---|
| | FeRAM[9] | PCM[9,10] | RRAM[9,10] | STT-MRAM[10] | SOT-MRAM (This work) |
| Speed | 10000 ps | 700 ps | 85 ps | <10000 ps | 400 ps |
| Endurance | $4 \times 10^6$ | $10^6$ - $10^9$ | $10^6$ - $10^9$ | $>10^{14}$ | $\sim 10^{15}$ ($> 10^{11}$ in our test) |
| Resistance Variation | 24.5% | 9.62% | 9.65% | 6.5% | < 4.8% |

| | Network Level | | | | |
|---|---|---|---|---|---|
| | BNN[5] | DNN[6] | SNN[7] | PNN | PBNN (This work) |
| Speed (Epochs when accuracy > 98%) | 72 | 30 | 23 | / | 4 |
| Required Bits Number | Binary | Multiple | Multiple | Multiple | Binary |
| Noise-Tolerant | Bad | Bad | Moderate | Good | Good |

Fig. 18. (a) Distributions of measured inference current compared to the target values in the last FCN layer. (b) The inference current corresponds to the input data 7.

Fig. 19. Classification results of 100 on-chip inferences.

Table. 1. Comparisons of device and network-level performance among different memristors and neural networks.